\documentclass[a4paper, 12pt]{article}
\usepackage[dvipsnames]{xcolor}  
\usepackage{aapreprint}
\usepackage{amsmath}
\usepackage{amssymb}
\usepackage{amsthm}
\usepackage{xparse}

\usepackage[english]{babel}
\usepackage{hyperref}
\usepackage{graphicx,wrapfig,float,slashed}
\usepackage{amsmath,amssymb,dsfont,epsfig,graphicx}
\usepackage{physics}
\usepackage{mathtools} 
\usepackage{multirow}
\usepackage{blkarray}
\usepackage{epstopdf}
\usepackage{ragged2e}
\usepackage{cancel}
\usepackage[toc]{appendix}
\usepackage[normalem]{ulem}
\usepackage{enumitem}
\usepackage{fontspec}
\usepackage{lmodern}
\allowdisplaybreaks
\usepackage{pifont}
\usepackage[style=base]{subcaption}
\usepackage{footnote}

\usepackage[compat=1.1.0]{tikz-feynman}
\usepackage{cleveref}

\newcommand{\nc}{\newcommand}
\nc{\ut}[2][]{\ensuremath{~\textrm{#2}^{#1}}}
\nc{\non}{\nonumber}
\nc{\hc}{\hbox {h.c.}}
\nc{\noi}{\noindent}
\nc{\barx}{\bar{x}}
\nc{\pbarn}{\;\hbox {pb}}
\nc{\fbarn}{\;\hbox {fb}}
\nc{\val}[3][]{\ensuremath{\qty(\frac{#2}{#3})^{#1}}}

\nc{\hsp}{\hspace{0.5cm}}
\nc{\lsp}{\hspace{1cm}}
\nc{\Lsp}{\hspace{2cm}}
\nc{\LLsp}{\lsp\lsp}
\nc{\lra}{\longrightarrow}
\nc{\p}{\prime}
\nc{\sgn}{\text{sgn}}
\nc{\ph}{\varphi}
\nc{\vmol}{v_{\text{M\o l}}}
\nc{\sun}{\odot}
\nc{\msun}{\ensuremath{\mathrm{M}_{\sun}}}

\newcommand{\beq}{\begin{equation}}
\newcommand{\eeq}{\end{equation}}
\newcommand{\bea}{\begin{eqnarray}}  
\newcommand{\eea}{\end{eqnarray}}
\newcommand{\baa}{\begin{array}}     \newcommand{\eaa}{\end{array}}
\newcommand{\bit}{\begin{itemize}}   \newcommand{\eit}{\end{itemize}}
\newcommand{\ben}{\begin{enumerate}} \newcommand{\een}{\end{enumerate}}
\newcommand{\bce}{\begin{center}}    \newcommand{\ece}{\end{center}}
\newcommand{\bpm}{\begin{pmatrix}}   \newcommand{\epm}{\end{pmatrix}}
\newcommand{\bvt}{\begin{verbatim}}  
\newcommand{\evt}{\end{verbatim}}

\newcommand{\cA}{\mathcal{A}}
\newcommand{\cB}{\mathcal{B}}

\newcommand{\cL}{\mathcal{L}}
\newcommand{\cM}{\mathcal{M}}

\newcommand{\cO}{\mathcal{O}}

\newcommand{\Gev}{\text{ GeV}}
\newcommand{\Mev}{\text{ MeV}}

\title{Resonant Enhancement for the transfer of baryon number from a CP-violating hidden sector}

\author[1]{ Can Kilic,}
\emailAdd{kilic@physics.utexas.edu}
\author[1]{\\ Sanjay Mathai}
\emailAdd{sanjaym@utexas.edu}
\affiliation[1]{Theory Group, Weinberg Institute for Theoretical Physics,\\ University of Texas at Austin, Austin, TX 78712, USA}

\abstract{We consider a scenario where the baryon asymmetry in the SM arises from a portal coupling to a CP-violating hidden sector, without any explicit violation of baryon or lepton number in either sector. While no net baryon number is generated, equal and opposite baryon numbers can be sequestered between the two sectors in this scenario. The observed asymmetry must be generated below the electroweak scale in order to avoid washout. By considering a benchmark model incorporating the baryon portal and a minimal hidden sector, we review the prospects of this scenario. When the asymmetry is generated via the decay of top quarks, generic input parameters for the model are sufficient. If however the asymmetry is to be generated via the decay of bottom quarks, a resonant enhancement is needed to sequester the baryon number with maximal efficiency. The scenario of mesogenesis assisted by CP-violation in a hidden sector falls into this latter category. We use analytical and numerical methods to quantify the amount of resonant enhancement that can be achieved, and we remark that the mesogenesis scenario can be discovered or ruled out in full if the relevant branching ratios can be improved by 2-3 orders of magnitude.}

\begin{document}
\preprint{UT-WI-18-2026}
\maketitle

\section{Introduction}
The Standard Model (SM) is an exceptionally successful theory describing the strong, weak and electromagnetic interactions. It has been tested over a wide range of energies and all of its predictions have been confirmed, even at the level of detailed precision measurements. Despite the success, the SM is not a complete theory of nature, and one of its most serious shortcomings is the failure to account for the observed matter-antimatter asymmetry in the universe. Cosmological observations show that there is an over-abundance of matter over antimatter in the universe quantified by the baryon to entropy ratio~\cite{ParticleDataGroup:2024cfk}, 
\begin{equation}
Y_B \equiv \frac{n_B-n_{\bar{B}}}{s}\approx 8.7\times 10^{-11},    
\end{equation}
where $n_B$ and $n_{\bar{B}}$ are the number densities of baryons and antibaryons respectively and $s$ is the total entropy density. Within the SM, this asymmetry cannot be achieved given symmetric initial conditions. Several models beyond the Standard Model (BSM) have been proposed to generate the asymmetry~\cite{Affleck:1984fy,Fukugita:1986hr,Cohen:1993nk,Dine:2003ax}.

In 1967, Sakharov formulated three necessary conditions for dynamically generating a matter-antimatter asymmetry \cite{Sakharov:1967dj}. These are summarized as,
\begin{enumerate}
    \item Baryon number (B) violation : If baryon number is conserved, then a universe that starts with $Y_B=0$ will remain at $Y_B=0$.
    \item C and CP violation (CPV): If C and CP are conserved, then a process with $\Delta B=+1$ would be accompanied by a $\Delta B=-1$ process with an equal rate.
    \item Out of equilibrium dynamics: In thermal equilibrium, processes that generate the asymmetry and the inverse process would occur at the same rate.
\end{enumerate}

The first condition can be relaxed if there is a hidden sector (HS), with baryon-number carrying particles, that is effectively decoupled from the SM at late times. If baryon number is transferred between the two sectors in the early universe, then equal and opposite abundances of $B$ can be sequestered, while the net $B$-number of the universe remains at zero.

While there are sources of CPV in the SM, in particular in the CKM matrix as well as in the $\theta$ parameter, constrained to be less than $\sim10^{-10}$\cite{Abel:2020pzs} by experiment, these are too small for generating the observed asymmetry. In principle, a HS which is entirely decoupled from the SM can contain an arbitrary amount of CPV. However, for a HS to help generate the SM asymmetry, it must be coupled to the SM. 

An attractive framework for coupling HSs, particularly in the case where the HS degrees of freedom are neutral under the SM gauge groups, to the SM is to add a ``portal'' interaction of the form $\frac{1}{M^\Delta}\cO_{\text{SM}}\cO_\text{HS}$ where $\cO_{\text{SM}}$ is a gauge invariant SM operator, $\cO_{\text{HS}}$ is an operator in the HS, and $M$ and $\Delta$ characterize the UV physics that the portal arises from.

The SM coupled to a HS via a portal can also satisfy the third Sakharov condition in a generic way, if particles in the SM can decay to the HS (or vice versa) out-of-equilibrium. Note also that as in many other mechanisms proposed to generate the matter antimatter asymmetry, weak decays and charge conservation are sufficient to also generate the necessary asymmetry in the charged leptons, with the lepton asymmetry being balanced by a background of anti-neutrinos. Thus, all the necessary ingredients for generating the asymmetry are present in such a setup.

With these observations in mind, let us define the question that will interest us in this paper. Specifically, {\bf we want to consider a HS coupled to the SM, with all the CPV existing in the HS, and with no explicit violation of $B$ (or $B-L$) number in either sector}. We then wish to study the necessary conditions for the observed baryon asymmetry (OBA) to be generated in this setup via baryon number being transferred between the two sectors at high energy, and then being locked in as equal and opposite amounts in each sector at low energy. 

An important requirement for this setup to work is that the asymmetry, once generated, must not be washed out by sphaleron processes. The baryon number transfer (via out-of-equilibrium decays) must therefore be active below the electroweak scale. This is in contrast to other commonly used ways of generating the asymmetry, such as leptogenesis~\cite{Fukugita:1986hr,Buchmuller:2004nz,Buchmuller:2005eh}, where $B-L$ is explicitly broken at high energy due to Majorana masses for heavy neutrinos, and thus the generated asymmetry cannot be washed out by sphaleron processes.

Since the baryon number transfer needs to be accomplished in a relatively narrow energy range, below the electroweak scale, but above the scale of BBN, the question arises whether the transfer can be made sufficiently efficient in order to generate the OBA as the universe cools. When an asymmetry is generated via decays of SM particles, in particular top or bottom quarks, CPV can only contribute via interference between tree and loop-level diagrams, making it suppressed by a loop factor. This may be sufficient when the branching ratio of the relevant decay channel can be large enough. However, when the branching ratio is bounded by existing constraints, this generic efficiency may not be enough. Even then, the loop factor suppression can be compensated by resonant enhancement. For the case of bottom quark decays, the scenario of mesogenesis~\cite{Elor:2018twp,Elor:2020tkc,Alonso-Alvarez:2021qfd,Elahi:2021jia,Elor:2022jxy,Elor:2025fcp} assisted by a CPV HS~\cite{Elor:2025fcp} belongs to the class of models we consider. Based on that study, given the limits on rare bottom decays, one can verify that a loop-suppressed efficiency for the transfer of baryon number is insufficient to generate the OBA. In this work we will show explicitly that an $\cO(1)$ efficiency can be achieved via resonant enhancement.

In order to quantify the level of resonant enhancement, we will set up a minimal model along the lines of ref.~\cite{Elor:2025fcp}. We will first consider the generation of  the asymmetry via top quark decays, with the conclusion that no resonant enhancement is needed. We will then turn our attention to the decays of bottom quarks. In this case, a resonant enhancement is indeed needed. We will analytically calculate the relevant Feynman diagrams and map out the region of parameter space where the OBA can be generated. We will also demonstrate via a Monte Carlo study that the resonant enhancement is generic and does not require a special structure in flavor space for the relevant input parameters.

The outline of the paper is as follows. In section \ref{sec:model}, we present our minimal benchmark model to make a resonantly enhanced transfer of baryon number possible. In section \ref{sec:applications} we consider the generation of the OBA from top and bottom quark decays. In section \ref{sec:experiment} we comment on the experimental probes for this setup. We conclude in section \ref{sec:conclusions}.

\section{Benchmark Model For the Hidden Sector}

\label{sec:model}

In this section, we introduce a minimal benchmark model containing only the necessary ingredients for a resonantly enhanced transfer of baryon number. The model we choose mirrors that of ref.~\cite{Elor:2025fcp} to make comparison easier.

Since our goal is to consider the transfer of baryon number between the two sectors via the decay of top and bottom quarks, a natural choice is to connect the SM to the HS via the ``baryon portal'' $\cO_{SM}=\bar{u}^c (\bar{d}^c \bar{d}^c)$ involving the right-handed up and down-type quarks. Here and in the rest of the paper we are using two-component spinor notation consistent with ref.~\cite{Dreiner:2008tw}. The brackets indicate a Lorentz singlet and a flavor-antisymmetric combination, and we are suppressing flavor indices for the moment. Our minimal model of the HS will contain only the essential ingredients for allowing a resonantly enhanced transfer of baryon number, though of course more complicated models of HSs can be considered to simultaneously address other outstanding problems, such as the microscopic nature of dark matter (DM).

A key element we need for resonant enhancement is to have at least two particles that have near degenerate masses and that act as the portal between the two sectors. The choice of the baryon portal calls for a fermionic portal particle, and thus we take these particles to be Dirac fermions $\Psi_a=(\psi_{a},\overline{\psi_{a}^c})$. The quantum numbers of $\Psi_a$ as well as other particles within the benchmark model are summarized in table~\ref{tab:newparticles}.

\begin{table}[!h]
\renewcommand{\arraystretch}{1.25}
  \setlength{\arrayrulewidth}{.25mm}
\centering
\small
\setlength{\tabcolsep}{0.18 em}
\begin{tabular}{|c || c | c| c | c | c | }
\hline
Field & Spin & SU$(3)_c$ & $U(1)_Y$ & $U(1)_B$  \\ \hline \hline
$\Phi$ & 0 & \textbf{3} & -2/3 & 2/3   \\ \hline
$\Psi_{a}$ & 1/2 & \textbf{1} & 0 & -1   \\ \hline
$S$ & 0 & \textbf{1}& 0 & +1  \\ \hline
$\chi$ & 1/2 & \textbf{1} & 0 & 0  \\ \hline
\end{tabular}
\vspace{0.1cm}
\caption{The quantum numbers of all BSM fields in the minimal model. The flavor index $a$ on $\Psi_a$ takes values $1,2$. }
\label{tab:newparticles}
\end{table}

Since $\cO_{SM}$ is a right-handed fermion with $B=+1$, we assign $B=-1$ to the $\Psi_a$ and we couple the right-handed components $\overline{\psi}_{a}^c$ to $\cO_{SM}$. Thus the two sectors interact via the dimension 6 operator
\begin{equation}
     \cL_{\text{eff}}\supset \sum_{ijka} \frac{c_{ijka}}{M^2} (\overline{\psi}^c_{a}\overline{u}_{i}^c) (\overline{d}^c_{j} \overline{d}^c_{k})   + \text{h.c.}.
\end{equation}
where the $i,j,k$ indices denote flavors of SM quarks, $a$ denotes the flavor of $\Psi$, the color indices of the quarks have an antisymmetric combination and $M$ is the mass scale suppressing the portal. This operator can arise from several possible UV completions. To be concrete, in the benchmark model we consider a heavy scalar mediator $\Phi$ which is a fundamental of $SU(3)_c$ and has hypercharge $-2/3$, with the couplings
\begin{equation}
     \cL \supset -\sum_{ij}\lambda_{ij}\Phi^* (\overline{d}_{i}^c\overline{d}_{j}^c)-\sum_{ia}\kappa_{ia}\Phi ( \overline{u}_{i}^c \overline{\psi}_{a}^c) + \text{h.c.}
\end{equation}
 In this UV completion, $M=M_{\Phi}$. Note that the color indices in the first term are contracted with an antisymmetric tensor, which requires that the flavor indices $(i,j)$ must also be in an antisymmetric combination. There are thus three independent $\lambda$ couplings, while there are six independent $\kappa$ couplings.
 
$\Phi$ can be resonantly produced at the LHC through the $\lambda$ coupling, as well as pair-produced through QCD, and decay either to $jj$ via the $\lambda$ couplings or to $j$+MET via the $\kappa$ couplings. The pair-production rate is independent of these couplings, and bounds from direct searches are satisfied for either decay mode as long as $M_{\Phi}\ge1.2$~TeV~\cite{ATLAS:2017jnp,CMS:2019zmd,ATLAS:2020syg}. For this mass, dijet resonance searches constrain $\lambda_{12}\lesssim 10^{-2}$\cite{Pascual-Dias:2020hxo}. We will not assume any particular flavor structure for the $\lambda_{ij}$ and $\kappa_{ia}$ couplings, and therefore we take all $\lambda$ couplings to have magnitudes of $O(10^{-2})$ or smaller. Direct searches for $\Phi$ do not place additional constraints on the $\kappa$ couplings.

$\Phi$ can give rise to flavor changing neutral currents (FCNC's) among down-type quarks at tree level, however the benchmark parameters listed above are consistent with existing constraints~\cite{Alonso-Alvarez:2021qfd}. The $\kappa$ couplings only contribute to FCNC's in the up-quarks at loop level, however due to the suppression from a loop factor as well as from $M_{\Phi}$, existing bounds~\cite{Golowich:2007ka} do not constrain the $\kappa$ couplings.

In order for $\Psi$'s to decay into the HS after they are produced from the SM, there must be kinematically accessible states in the  HS to which they can decay. Since $\Psi$ carries baryon number, at least one of the HS particles must carry baryon number as well. For minimality, we introduce a Majorana fermion $\chi$ and a scalar $S$ with $B(S)=+1$, with the added HS interaction
\begin{equation}
    \cL \supset -\sum_a y_a(\psi_a \chi) S +  {\rm h.c.}.
\end{equation}
This completes the benchmark model. With this interaction as well as those of $\Phi$, the HS does not have any additional global $U(1)$ symmetries in addition to baryon number. In order to avoid proton decay, and to keep both $\chi$ and $S$ stable, we choose the masses $(m_p,m_\chi,m_S)$ to be consistent with the sides of a triangle, thereby making sure that none can decay to the other two. We assume $m_{\Psi}>m_{\chi}+m_S$ such that the decay of $\Psi$ into HS states is kinematically allowed. Since $\chi$ and $S$ are stable and neutral under the SM, they are potentially good DM candidates. In this study we only focus on the transfer of baryon number, but see ref.~\cite{Elor:2025fcp} for a more ambitious attempt at also reproducing the correct DM abundance.

The BSM couplings of the benchmark model contain CP phases that can be $O(1)$. In particular, there are three phases in $\lambda_{ij}$, six phases in $\kappa_{ia}$ and two phases in $y_a$. Not all these phases are physically observable. We define three invariants (one for each up-type quark flavor $k$)
\bea
\Delta_k \equiv \text{Im}(\kappa_{k1}\kappa_{k2}^*\,y_1\, y_2^*),
\label{eq:defdelta}
\eea
which directly appear in our calculations in the rest of the paper and which cannot be removed via rephasing fields.
As we will see below, the $\lambda_{ij}$ phases do not contribute to the generation of the asymmetry in this model.
Let us now turn our attention to the transfer of baryon number between the two sectors.

With these couplings, quarks, if kinematically allowed, can now decay into states where one of the final states is a portal fermion $\Psi_a$, with $\Psi_a$ decaying predominantly into the HS. If sufficient CPV is present, these processes can therefore sequester baryon number between the visible and HSs, generating the OBA. 

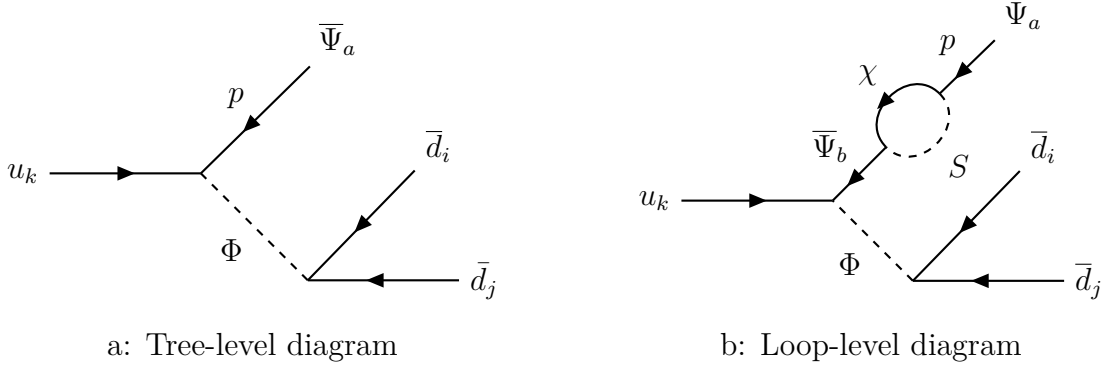
\begin{figure}[h!]
\centering
\begin{subfigure}[t]{0.45\textwidth}
   \centering
   \begin{tikzpicture}
   \begin{feynman}    
       \vertex (d1) ; 
       \vertex [right=2cm of d1] (v1);   
       \vertex [above right=2cm of v1] (d11) {$\overline{d}_i$};
       \vertex[ right=2cm of v1] (u) {$\bar{d}_j$};     
       \vertex[above left=2cm of v1] (v2);
       \vertex[above right=2cm of v2] (p) {${\overline{\Psi}_a}$};
       \vertex[ left=2cm of v2] (d2) {$u_k$};
        
       \diagram* {       
           (d11) -- [fermion, thick] (v1) -- [anti fermion, thick] (u),
           (v1) -- [scalar, thick, edge label=$\Phi$] (v2),
           (d2) -- [fermion, thick] (v2) -- [anti fermion, thick, edge label=$p$] (p),
       };   
   \end{feynman}
   \end{tikzpicture}
   \caption{Tree-level diagram}
   \label{fig:treeu}
\end{subfigure}
\hfill
\begin{subfigure}[t]{0.45\textwidth}
   \centering
   \begin{tikzpicture}
   \begin{feynman}    
       \vertex (d1) ; 
       \vertex [right=2cm of d1] (v1);        
       \vertex[ right=2cm of v1] (u) {$\overline{d}_j$};  
       \vertex [above right=2cm of v1] (d11) {$\overline{d}_i$};
       \vertex[above left=1.5cm of v1] (v2);
       \vertex[above right=1cm of v2] (v3);
       \vertex[above right=1cm of v3] (v4);
       \vertex[above right=1cm of v4] (p) {${\overline{\Psi}_a}$};
       \vertex[ left=2cm of v2] (d2) {$u_k$};
        
       \diagram* {       
           (d11) -- [fermion, thick] (v1) -- [anti fermion, thick] (u),
           (v1) -- [scalar, thick, edge label=$\Phi$] (v2),
           (d2) -- [fermion, thick] (v2) -- [anti fermion, thick, edge label=$\overline{\Psi}_{b}$] (v3) 
                 -- [anti fermion, half left, thick, edge label=$\chi$] (v4) 
                 -- [anti fermion, thick, edge label=$p$] (p),
           (v4) -- [scalar, half left, edge label=$S$, black, thick] (v3),
       };   
   \end{feynman}
   \end{tikzpicture}
   \caption{Loop-level diagram}
   \label{fig:loopu}
\end{subfigure}
\vspace{0.2cm}
\caption{Feynman diagrams that induce CPV in up-type quark decays.}
\label{fig:psi-production-u}
\end{figure}

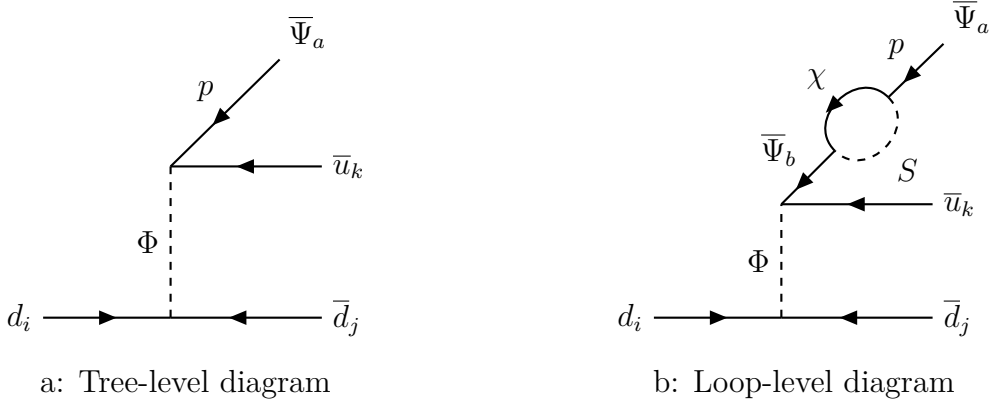
\begin{figure}[h!]
\centering
\begin{subfigure}[t]{0.45\textwidth}
   \centering
   \begin{tikzpicture}
   \begin{feynman}    
       \vertex (d1) {$d_i$}; 
       \vertex [right=2cm of d1] (v1);        
       \vertex[ right=2cm of v1] (u) {$\overline{d}_j$};     
       \vertex[above=2cm of v1] (v2);
       \vertex[above right=2cm of v2] (p) {${\overline{\Psi}_a}$};
       \vertex[ right=2cm of v2] (d2) {$\overline{u}_k$};
        
       \diagram* {       
           (d1) -- [fermion, thick] (v1) -- [anti fermion, thick] (u),
           (v1) -- [scalar, thick, edge label=$\Phi$] (v2),
           (d2) -- [fermion, thick] (v2) -- [anti fermion, thick, edge label=$p$] (p),
       };   
   \end{feynman}
   \end{tikzpicture}
   \caption{Tree-level diagram}
   \label{fig:treed}
\end{subfigure}
\hfill
\begin{subfigure}[t]{0.45\textwidth}
   \centering
   \begin{tikzpicture}
   \begin{feynman}    
       \vertex (d1) {$d_i$}; 
       \vertex [right=2cm of d1] (v1);        
       \vertex[ right=2cm of v1] (u) {$\overline{d}_j$};     
       \vertex[above=1.5cm of v1] (v2);
       \vertex[above right=1cm of v2] (v3);
       \vertex[above right=1cm of v3] (v4);
       \vertex[above right=1cm of v4] (p) {${\overline{\Psi}_a}$};
       \vertex[ right=2cm of v2] (d2) {$\overline{u}_k$};
        
       \diagram* {       
           (d1) -- [fermion, thick] (v1) -- [anti fermion, thick] (u),
           (v1) -- [scalar, thick, edge label=$\Phi$] (v2),
           (d2) -- [fermion, thick] (v2) -- [anti fermion, thick, edge label=${\overline{\Psi}_b}$] (v3) 
                 -- [anti fermion, half left, thick, edge label=$\chi$] (v4) 
                 -- [anti fermion, thick, edge label=$p$] (p),
           (v4) -- [scalar, half left, edge label=$S$, black, thick] (v3),
       };   
   \end{feynman}
   \end{tikzpicture}
   \caption{Loop-level diagram}
   \label{fig:loopd}
\end{subfigure}
\vspace{0.2cm}
\caption{Feynman diagrams that induce CPV in down-type quark decays.}
\label{fig:psi-production}
\end{figure}

CPV in decays appears as the result of interference  between tree and loop level diagrams. In our model, at the parton level, the diagrams reponsible for the CPV are shown in Fig.~\ref{fig:psi-production-u} for up-type quark decays, and Fig.~\ref{fig:psi-production} for down-type quark decays.

In the decay of a quark $q_i$ with number density $n_i$ to lighter quarks and $\Psi_a$, the total baryon number density that can be sequestered between the two sectors is proportional to
\begin{equation}
    n_i \sum_{a,j,k}\left(\text{Br}(\overline{q}_i\rightarrow \Psi_aq_jq_k)- \text{Br}({q}_i\rightarrow \overline{\Psi}_a\overline{q}_j\overline{q}_k) \right).
\end{equation}
Focusing on only one pair of SM quarks in the final state at a time, one can define a convenient dimensionless measure of the effectiveness of the baryon-number transfer, given by
\begin{equation}
    A_{CP}^{\{i,j,k,a\}} \equiv 
 \frac{
 \Gamma (\overline{q}_i\rightarrow \Psi_a {q}_j{q}_k) - \Gamma (q_i  \rightarrow \overline{\Psi}_a \overline{q}_j \overline{q}_k )}{ \Gamma (\overline{q}_i\rightarrow {\Psi}_a {q}_j{q}_k) + \Gamma (q_i  \rightarrow \overline{\Psi}_a \overline{q}_j \overline{q}_k )}.
 \label{eq:ACPdark}
\end{equation}
From here onwards, we will suppress the indices on $A_{CP}$, unless they are explicitly needed for context. For a specific set of quarks in the final state, the baryon asymmetry that is generated can be written as 
\begin{equation}
    \sum_{a} n_i\, {\rm Br}_{{\rm BSM},i,a}\,A_{CP}^{i,a},
\end{equation}
where ${\rm Br}_{{\rm BSM},i,a}$ is the BSM branching ratio of $q_i$ to $\Psi_a$. Thus the $A_{CP}$ are a measure of the efficiency of baryon number transfer. Here we explicitly sum over the $a$ index, as the total baryon number transfer will involve both $\Psi$ flavors in the final state, and we need to keep track of how they add up. For generic values of the $\lambda$ and $\kappa$ couplings, we expect the two $A_{CP}^{i,a}$ with $a=1,2$ to have similar magnitudes. As we will show soon, in our benchmark model, they also have the same sign, so there is no cancellation of the transferred baryon number when adding over final states.

For typical scenarios without a resonant enhancement, the $A_{CP}$ which comes from the interference between tree level and loop level diagrams is suppressed by a loop factor $\frac{g^2}{16\pi^2}\sim 10^{-3}$. Our goal is to show that resonant enhancement can bring $A_{CP}$ up to an $\cO(1)$ value, which as we will see in the next section is essential to generating the OBA if this is to be accomplished through the decay of bottom quarks.

\section{Implementation}
\label{sec:applications}

In this section, we identify the region of parameter space that gives rise to the OBA in two scenarios, with top and bottom quarks decaying through the portal interaction. 
This sets the target range for the product of the top/bottom branching fraction through the portal, and the value of $A_{\rm CP}$. 

\subsection{Top decays}

In this section, we consider generating the baryon asymmetry from decays of the top quark. We remind the reader that only decays that occur after the sphaleron processes are turned off can contribute to generating the asymmetry, namely at $T\lesssim 100 \Gev$. Top quarks are non-relativistic at this point, and we need to set up a Boltzmann equation in order to correctly estimate the generated asymmetry. The conclusion of this section is going to be that even relatively inefficient transfer of baryon number can generate the OBA, the couplings being many orders of magnitude away from experimental constraints. This being the case, it will suffice to drop order one factors in our calculations on occasion, as they would not affect the overall conclusion. In the next section we will find that the constraints are much tighter when generating the asymmetry through the decay of bottom quarks and therefore enhancing the efficiency of the transfer is essential.

We define the yield $Y_i\equiv \frac{n_i}{s}$ where $n_i$ is the number density of particle $i$ and $s$ is the total entropy density. Moreover, we also define the dimensionless quantity $x=\frac{m_t}{T}$. For simplicity, we will start our calculation when $T=100$~GeV and the sphaleron processes are absent, and we will integrate down to 10~GeV, after which point top quarks are essentially absent from the plasma. As mentioned above, this is one of our simplifying assumptions which may lead to an order one inaccuracy, but will not change the overall conclusion. Since we are only tracking the asymmetry, let us furthermore focus only on the SM process $b+W\leftrightarrow t$ as well as the CP conjugate processes as the dominant top-number changing process. Top-number preserving processes such as $gg\to t\overline{t}$ which have both a top and an anti-top in the final state are in any case ineffective at these lower temperatures, and their main effect is in setting the equilibrium density for tops. 
With this setup the Boltzmann equation for the number density of tops can be put into the simple form
\begin{equation}
    \frac{dY_t}{dx} = -\frac{\Gamma_t}{H(x)x}\bigl(Y_t-Y_t^{\rm eq}\bigr)
                      -\frac{\Gamma_t^{\rm HS}}{H(x)x}\,Y_t,
\end{equation}
where $\Gamma_t^{\text{HS}}$ is the width of the decay of top quark into $\overline{\Psi}+$SM. Since we are only using a benchmark value for the relevant $\lambda$ and $\kappa$ couplings, we do not put in an additional factor for the number of distinct final states.

\begin{figure}
    \centering
    \includegraphics[width=0.75\linewidth]{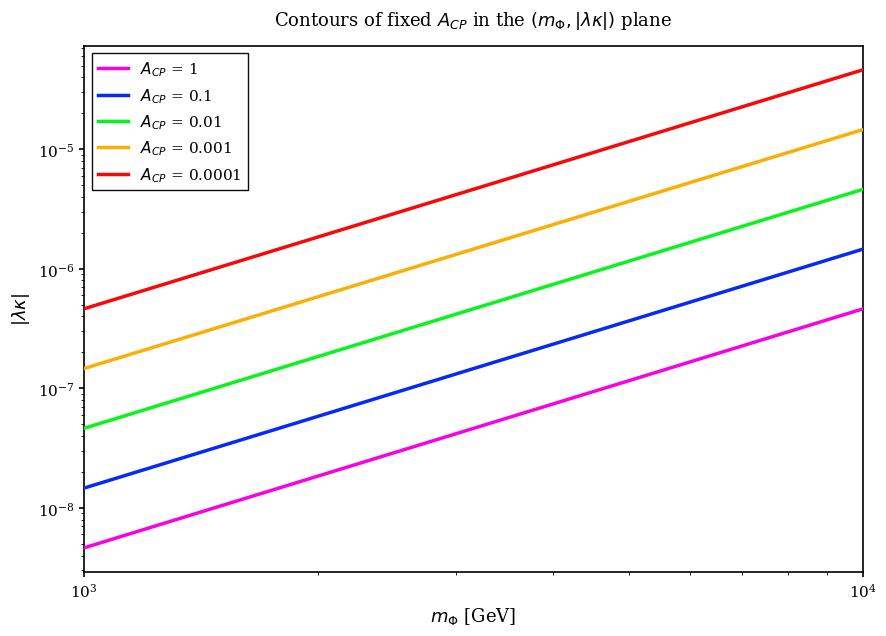}
    \caption{Values of $|\lambda\kappa|$ and $m_\Phi$ that generate the OBA from top quark decays.}
    \label{fig:lamcurves}
\end{figure}

In the range $x\in [1.73,17.3]$, we have $\Gamma_t\gg H(x)\gg \Gamma_t^\text{HS}$, such that the tops are kept near their equilibrium density and we can apply a quasi-static approximation, 
\begin{equation}
  Y_t(x)
  = \frac{\Gamma_t}{\Gamma_t+\Gamma_t^{\rm HS}}\,Y_t^{\rm eq}(x),
  \label{eq:YQS}
\end{equation}
where 
\begin{equation}
  Y_t^{\rm eq}(x) \approx 0.0081\;x^{3/2}e^{-x}.
  \label{eq:Yteq}
\end{equation}  
Then, the baryon asymmetry is accumulated as
\begin{equation}
Y_B \approx \int_{x_0}^{x_f} \frac{dY_B}{dx}\,dx
\approx \int_{x_0}^{x_f} A_{CP}\,\frac{\Gamma_t^{\rm HS}}{H(x)x}\,Y_t(x)\,dx
  \label{eq:dYB}
\end{equation}
The OBA is given by $Y_B^{\rm obs} = 8.7 \times 10^{-11}$, which leads to the preferred value of 
\begin{equation}
    A_{\rm CP}\times\Gamma_t^{\rm HS} \approx 1.6 \times 10^{-22} \Gev .
    \label{eq:topBR}
\end{equation}
In our model, the parametric dependence of $\Gamma_t^{\rm HS}$ is given by
\begin{equation}
    \Gamma_t^{\rm HS} \approx 0.5 \times 10^{-4} \times |\lambda|^2 |\kappa|^2 \frac{m_t^5}{m_\Phi^4},
\end{equation}
where the numerical factor in front arises from integrating over 3-body phase space. This leads to lines of constant $\lambda\kappa / M_{\Phi}^{2}$ for a given value of $A_{CP}$. This is shown in figure~\ref{fig:lamcurves}. Note that based on equation~\ref{eq:topBR}, the BSM branching ratio of the top would be of $\cO(10^{-19})$ for $A_{CP}\sim 10^{-3}$, which is unconstrained by collider searches.

As previously mentioned. the OBA can be generated even for very small values of the couplings, and thus a resonant enhancement of $A_{CP}$ is unnecessary. 

\subsection{Bottom decays}

As we have seen in the previous section, if our benchmark HS is coupled to the SM at $T\sim100$~GeV and the relevant couplings are in the appropriate range, the OBA can be generated in a straightforward manner, without the need for resonant enhancement. A downside of this setup is that the necessary couplings are too small for the new physics to be probed at current or upcoming experiments, unless $M_{\Phi}$ happens to be only slightly heavier than the current LHC bounds.

One can however imagine scenarios where the portal only couples to the lighter up-type quarks, or the SM is never in equilibrium at temperatures where top quarks are abundant in the plasma. The scenario of mesogenesis~\cite{Elor:2018twp,Elor:2020tkc,Alonso-Alvarez:2021qfd,Elahi:2021jia,Elor:2022jxy,Elor:2025fcp} falls into this latter category. In mesogenesis, it is assumed that after inflation the universe reheats to a low temperature below the QCD scale, and the first particles to be created (non-thermally) are bottom quarks via the decay of a reheaton field $\varphi$ with $m_\varphi\sim \cO(10-100\Gev)$. The bottom quarks hadronize first and then decay, establishing the SM plasma. The OBA is generated through rare CPV decays of bottom quarks.

There are significant constraints on the mesogenesis scenario. Since rare decays of bottom quarks have been searched for at the BaBar and Belle experiments~\cite{Belle:2021gmc,BaBar:2023rer}, the bounds on non-SM decays are at the order of $10^{-5}$. This makes generating the OBA a challenge. In fact, the original mesogenesis proposal~\cite{Elor:2018twp} which contains no new degrees of freedom in addition to the SM is no longer viable, as the limited amount of CPV in that setup combined with these low branching ratios is not sufficient for generating the OBA. As a result, more recent variations of the mesogenesis mechanism involve a HS as a more efficient source of CPV. 

In particular, ref.~\cite{Elor:2025fcp} conducts a detailed phenomnological study of mesogenesis coupled to the same minimal HS on which we have based our benchmark model. In the model of ref.~\cite{Elor:2025fcp}, the goal is not only to generate the OBA but to also identify the states in the HS as DM candidates with the correct relic abundance. If the efficiency by which the HS states annihilate is somewhat low, the DM retains a symmetric as well as an asymmetric component after freezeout, which allows for flexibility in the choice of masses for these states, allowing them to be light enough for bottom quarks to decay into them.

After setting up Boltzmann equations for both the baryon asymmetry and the DM abundance, it is shown in ref.~\cite{Elor:2025fcp} that 
\begin{equation}
\label{eq:master}
\frac{Y_\cB}{Y_\cB^{\rm obs}} \simeq \frac{\rm{Br}_\cM}{1.3\times 10^{-8}}\left(\frac{T_R}{60\Mev}\right)\frac{2 m_{\mathcal{M}}}{m_\varphi}\sum_a \bigl|A_{\rm CP}^{b,a}|,
\end{equation}
where $A_{CP}^{b,a}$ needs to be defined in terms of the decays of a bottom meson ${\mathcal M}$ to a SM baryon $\mathcal{B}_{\mathrm{SM}}$ (and possibly additional particles)
\begin{equation}
\label{eq:ACPmes}
A_{\rm CP}^{b,a}\equiv\frac{\Gamma(\overline{\mathcal{M}}\to\overline\Psi_{a}\overline{\mathcal{B}}_{\mathrm{SM}}) - \Gamma(\mathcal{M}\to\Psi_{a}\mathcal{B}_{\mathrm{SM}})}{\Gamma(\overline{\mathcal{M}}\to\overline\Psi_{a}\overline{\mathcal{B}}_{\mathrm{SM}}) + \Gamma(\mathcal{M}\to\Psi_{a}\mathcal{B}_{\mathrm{SM}})}.
\end{equation}
Combining Eq.~\eqref{eq:master} with the current direct-search bound $\text{Br}_\cM\lesssim 10^{-5}$ from BaBar/Belle and with $m_{\varphi}>2m_b\sim10$~GeV requires $A_{CP}^{b,a}$ to be above the generic loop-level value. Pushing the bound to the Belle-II projected reach $\text{Br}_\cM\lesssim 10^{-7}$ requires $|A_{CP}^{b,a}|\gtrsim 0.1$. Below $\text{Br}_\cM\sim 1.3\times 10^{-8}$, even $A_{CP}^{b,a}\sim1$ is insufficient to generate the OBA. 

In simple terms then, this represents a ``final stand'' for the mesogenesis mechanism. If the bounds on rare bottom meson decays can be pushed down below $1.3\times 10^{-8}$, then even maximally efficient CPV in a HS fails to generate the OBA, and decays of a non-thermal population of bottom mesons at temperatures below the QCD scales can be ruled out entirely as a possible source of the OBA. From a different perspective, this can be interpreted as an ultimate prediction of mesogenesis of the existence of rare bottom meson decay channels that are close to the projected limits. In ref.~\cite{Elor:2025fcp}, it was assumed that $A_{CP}^{b,a}\sim 1$ could be achieved via coupling to a HS. Below, we confirm this assumption, confirming that mesogenesis with a CPV HS is still a viable solution to the baryon asymmetry puzzle, which can be fully tested and ruled out with further improvements in the sensitivity of searches for rare bottom decay channels.

\begin{figure}
    \centering
    \includegraphics[width=0.75\linewidth]{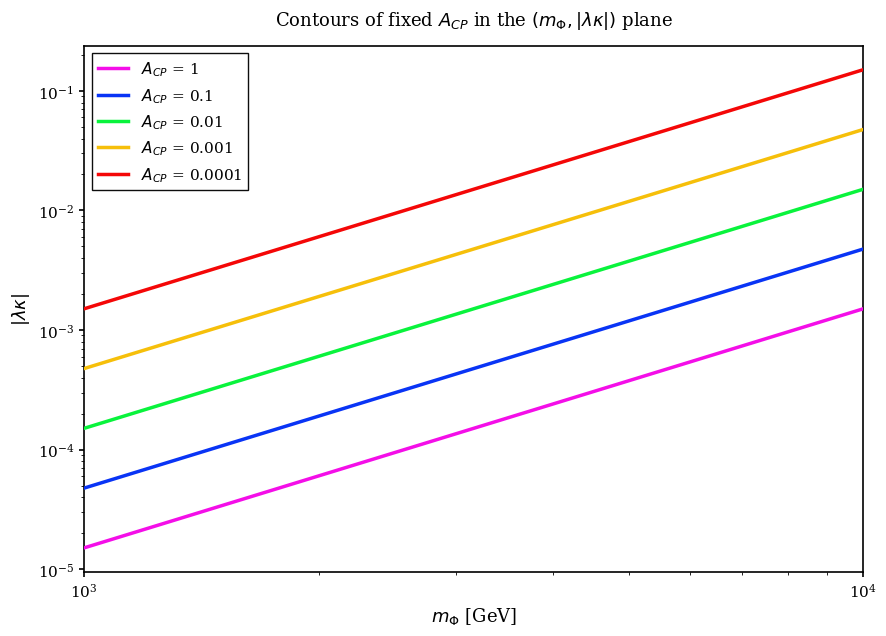}
    \caption{Values of $|\lambda\kappa|$ and $m_\Phi$ that generate the OBA from bottom quark decays.}
    \label{fig:meso}
\end{figure}

In Fig.~\ref{fig:meso}, we show the relation between the coupling constants and $m_\Phi$ that generates the OBA for $m_{\varphi}=2m_{\mathcal M}$ and $T_R=60$~MeV. 

\subsection{Quantifying the resonant enhancement}
\label{resenhCPV}
We now show that $A_{CP}$ can naturally be $\cO(1)$ in our minimal model. 
We write the decay rate as $\Gamma_{\rm HS} = \int d{\rm LIPS}_3 |\mathcal{A}_t +  \mathcal{A}_l|^2$ where the scattering amplitude $\mathcal{A}$ has been separated into tree and 1-loop level contributions and $d{\rm LIPS}_3$ is the three-body phase space volume element. Since the tree level contribution dominates the total rate, we may approximate:
\bea
A_{CP}^i \approx \frac{\text{Im} [\cA_t \cA_l^*] }{|\cA_t|^2 } \,.
  \label{eq:approxACP}
\eea

Figures \ref{fig:psi-production-u} and \ref{fig:psi-production} show the contributing diagrams for the decay of top and bottom quarks, respectively. 
Note that everything is common to the two diagrams except for the leg carrying the momentum $p$ shown in the figure. As shown in the appendix, the CPV in the HS arises from the propagator mixing among  $\Psi_a$. Following \cite{Pilaftsis:1997jf,Pilaftsis:2003gt},
the absorptive part of the 1PI corrections to the 
 $\Psi_a$-$\Psi_b$ propagator can be written as
\begin{equation}
    \Sigma_{ab}^\text{abs} = -\left[\frac{y_a(y_b)^*}{8\pi^2 p^2}K(p^2,m_\chi^2,m_S^2)^\frac{1}{2}\right] \slashed{p}  \qquad a\neq b \,, 
     \label{PI}
\end{equation}    
 where $p$ is the momentum running along the propagator,
and $ K(x,y,z) = x^2 + y^2 + z^2 -2xy - 2yz - 2zx$
is the Källen function.  In such process, CPV is resonantly enhanced when the masses of $\Psi$'s are nearly degenerate. When $m_{\Psi_1}=m_{\Psi_2}$, the internal propagator in the loop diagram goes on shell and is only suppressed by finite width effects \cite{Pilaftsis:1997jf,Pilaftsis:2003gt}. Generically, $\Gamma_{\Psi_1}\approx \Gamma_{\Psi_2}\equiv \Gamma$ and the resonant enhancement is most efficient when $\Gamma\sim \Delta m_{\Psi} \equiv m_{\Psi_1} - m_{\Psi_2}$. Then, as we show in the appendix
\begin{align}
A_{CP}^1 \approx \frac{\Delta_k}{4\pi|\kappa_{k1}^2|}\frac{m_{\Psi}\Delta m_{\Psi}}{(\Delta m_{\Psi})^2+\Gamma^2},
    \label{eq:ACPres}
\end{align}
up to higher-loop corrections. For $A_{CP}^2$, we exchange $1 \leftrightarrow 2$ in equation~\ref{eq:ACPres}, which changes the sign of $\Delta_k$ but also of $\Delta m_{\Psi}$, thus the two $A_{CP}^q$ are guaranteed to always have the same sign.

Strictly speaking, the on-shell renormalization scheme we have used here is only valid when $\Gamma \ll \Delta m_{\Psi}$, and needs to be replaced by a density matrix method when $\Gamma \sim \Delta m_{\Psi}$~\cite{Garny:2011hg}. Here we use this simpler form, however we have verified that in the resonant regime, our approach is within a factor of 2 of the result from the density matrix method.

In our benchmark model, since $A_{CP}^a \rightarrow 0$ as $\Delta m_{\Psi}\to 0$ with  $\Gamma$ fixed,  to achieve sizable dark CPV, $\Delta m_\Psi$ must not be significantly smaller than $\Gamma$ (in that limit the decays to $\Psi_{1,2}$ are no longer distinguishable). An $\mathcal{O}(1)$ $A_{CP}$ is obtained for example with $|\kappa|\sim \cO(1-10^{-2})$,  $|y|\sim \mathcal{O}(10^{-3})$, phases of $\mathcal{O}(1)$, and with $\Gamma / m_{\Psi}\sim \mathcal{O}(10^{-6})$. 
\begin{figure}
    \centering
\includegraphics[width=0.75\textwidth]{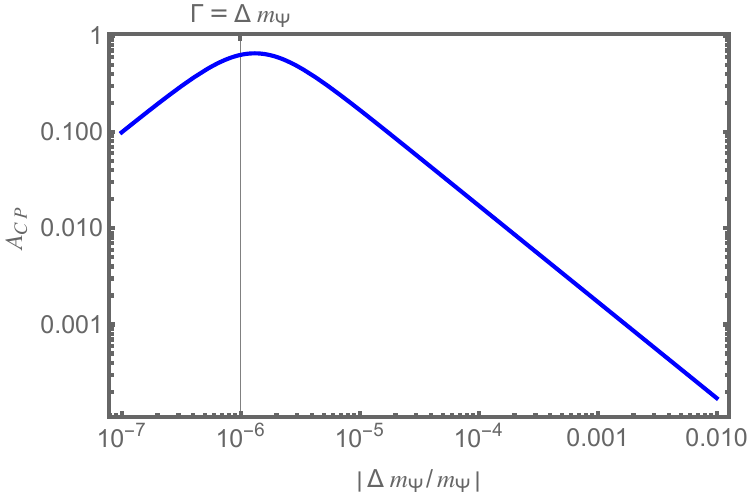}
\vspace{0.5cm}
    \caption{Numerical value of $A_{CP}$ for benchmark values $\kappa_{k1}=\kappa_{k2}^*=\frac{1}{\sqrt{2}}e^{i\frac{\pi}{4}},y_{1}=y_2=0.005,
    m_{\Psi}=4\text{ GeV},m_\chi=m_S=1\text{ GeV}$.}
    \label{CPV}
    \vspace{0.5cm}
\end{figure}
We show in Fig.~\ref{CPV} the dependence of $A_{CP}$ on  $\Delta m_{\Psi}/m_{\Psi}$ for fixed $\Gamma$, reaching a maximum value $A_{CP}\approx 0.7$. This figure is calculated using the exact definition of $A_{CP}$, without making any approximations. As expected, a resonant enhancement occurs close to the degenerate point where $\Delta m_{\Psi}\sim \Gamma$. In order to address whether the a tuned choice of phases or magnitudes of the couplings is necessary to achieve this kind of resonant enhancement, we describe below a numerical Monte Carlo study. 

\subsection{Monte Carlo study of the model parameters}

In our study we are operating with the assumption that the complex CP phases in the portal and dark sector couplings are random and do not have a specific structure in flavor space. In this section we confirm based on a numerical study that an $A_{CP}$ of order one can be generated without fine tuning with this assumption, in the relevant range of the magnitudes of the model couplings.  Since the amount of CPV depends on a large number of unknown variables, the simplest approach is to perform a Monte Carlo study over the parameter space.

\begin{figure}[!h]
    \centering
    \includegraphics[width=0.55\linewidth]{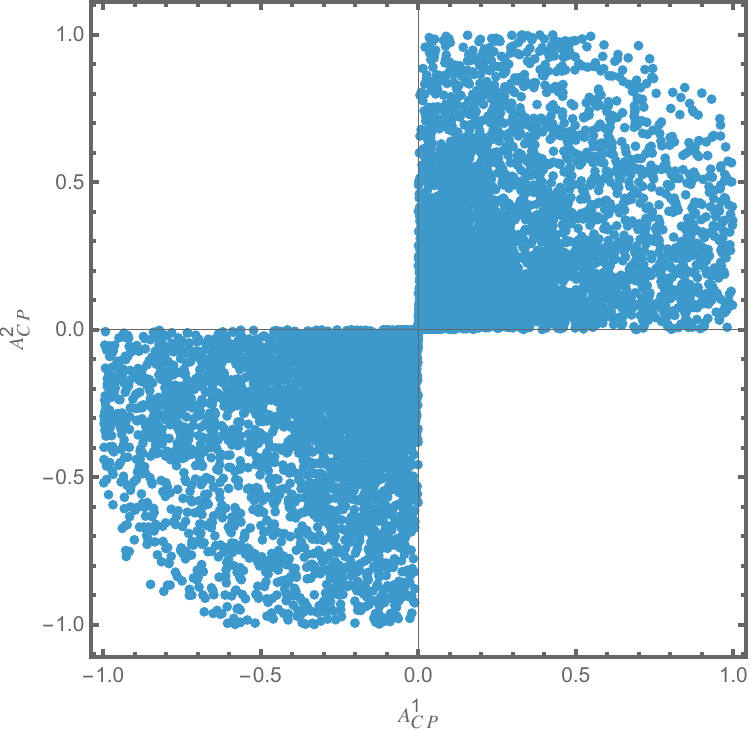}
    \caption{Scatter plot for $A_{CP}^{1},A_{CP}^{2}$.}
    \label{mcplot}
\end{figure}

\begin{figure}[!h]
     \centering
     \includegraphics[width=0.55\linewidth]{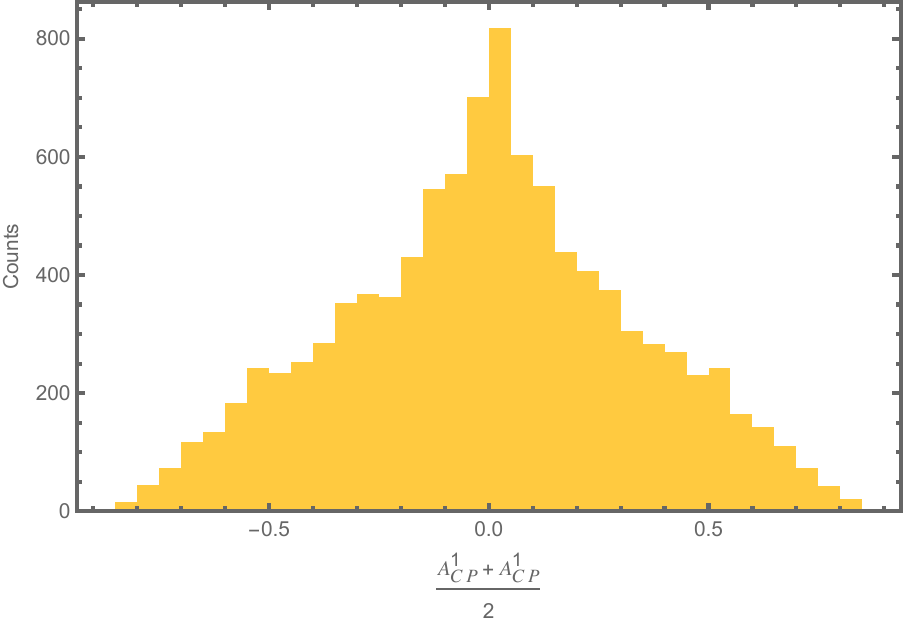}
     \caption{Histogram showing the population of $\frac{A_{CP}^{1}+A_{CP}^{2}}{2}$ in the Monte-Carlo simulation.}
     \label{hist}
 \end{figure}

The model parameters for the Monte Carlo study are chosen as follows:
 \begin{itemize}
     \item We fix $m_{\Psi}=4$ GeV with bottom quark decays in mind, and $m_\chi=m_{S}=1$ GeV such that the particles in the loop can go on shell.
     \item All couplings are chosen in the perturbative regime, and consistent with the experimental constraints discussed in the previous sections. We sample the magnitude $|y_a|$ for $a=1,2$ from a Gaussian distribution centered at $5\times 10^{-3}$ with a width of $3.5\times 10^{-3}$. Similarly, the magnitudes $|\kappa_{ka}|$ for $a=1,2$ and $k=1,2,3$ are sampled from a Gaussian distribution centered at $5\times 10^{-2}$ with a width $3.5\times 10^{-2}$. $A_{CP}^a$ has no direct dependence on $\lambda_{ij}$.
     \item We choose $\Delta m_\Psi = 2\times 10^{-6}$ GeV. This choice is to make sure that the scan includes the region where $\Delta m_{\Psi}\sim \Gamma$, where the CPV is most efficient, with the magnitudes of the couplings listed above. 
     \item We sample the phases of all couplings $|y_a|$ and $|\kappa_{ka}|$ uniformly in $[0,2\pi]$.
 \end{itemize}
   The population we obtain for $(A_{CP}^{1},A_{CP}^{2})$ is shown in Fig.~\ref{mcplot}.
We find that there is an $\mathcal{O}(1)$ probability of obtaining $A_{CP}^a$ above $0.5$. To better visualize the distribution underlying  the scatter plot in Fig.~\ref{mcplot}, we also present a one dimensional  histogram showing the average of $A_{CP}^{1}$ and $A_{CP}^{2}$ (see Fig.~\ref{hist}). Quantitatively, approximately $10\%$ of the points yields an average of $A_{CP}^{1}$ and $A_{CP}^{2}$ greater than or equal to $0.5$. Furthermore, the average $A_{CP}$ value is at or below the generic (non-resonant value) only a tiny fraction of the time ${\mathcal O}(10^{-4}-10^{-5})$.

\section{Experimental signatures}
\label{sec:experiment}

In this section we review potential experimental probes for our benchmark model. The relevant experimental searches can be divided into two categories of direct searches for the BSM particles, and precision measurements of observables that are affected by the existence of those particles, in particular searches for a neutron electric dipole moment.

\subsection{Direct searches }

As we have seen in the previous section, when the OBA is generated via top quark decays, the couplings that are involved are very small, and the only promising channel for directly producing the new physics is via the pair production of $\Phi$ through QCD. As discussed in section~\ref{sec:model}, prompt decays of $\Phi$ are only constrained in the region $M_{\Phi}<1.2$~TeV via QCD pair production, and $\lambda_{ds}>10^{-2}$ at this mass and above via resonant production. Despite the small couplings however, the $\Phi$ decays promptly on collider times scales: for $M_{\Phi}=1.2$~TeV, $c\tau\leq 0.4$~nm for $\lambda$ and $\kappa$ couplings that lie on any of the curves of figure~\ref{fig:lamcurves}. Therefore R-hadron searches~\cite{Barnard:2015rba} do not place additional constraints on the parameter space. 

When the asymmetry transfer is accomplished via bottom quark decays, the couplings involved are larger, allowing new physics to be constrained via searches for rare bottom decays. This is in fact the reason why $A_{CP}\sim 1$ is needed for generating the OBA. For different flavor structures in the baryon portal, table 1 of Ref.~\cite{Elor:2025fcp} lists the most promising decay channels for direct searches. 

\subsection{EDM constraints}
\begin{figure}[h!]
    \centering
    
    \begin{tikzpicture}[line width=1pt]
    \begin{feynman}
  \vertex (c1) at ({2*cos(210)},{2*sin(210)});   
  \vertex (c2) at ({2*cos(90)},{2*sin(90)});     
  \vertex (c3) at ({2*cos(330)},{2*sin(330)});   

  \vertex (g1) at ({4*cos(210)},{4*sin(210)});   
  \vertex (g2) at ({4*cos(90)},{4*sin(90)});    
  \vertex (g3) at ({4*cos(330)},{4*sin(330)});   

  \vertex (a1) at ({2*cos(180)},{2*sin(180)});   
  \vertex (a2) at ({1*cos(180)},{1*sin(180)});   
  \vertex (a3) at ({1*cos(0)},{1*sin(0)});       
  \vertex (a4) at ({2*cos(0)},{2*sin(0)});

  \draw[scalar] (2,0) arc[start angle = 0 , end angle = 180, radius = 2cm]  ;
  \draw (-2,0) arc[start angle = 180 , end angle = 360, radius = 2cm];
  \draw[gluon] (g1) -- (c1);
  \draw[gluon] (g2) -- (c2);
  \draw[gluon] (g3) -- (c3);
  \draw (a1) -- (a2);
  \draw (a3) -- (a4);
  \draw[scalar] (1,0) arc[start angle = 0 , end angle = 180, radius = 1cm];
  \draw (-1,0) arc[start angle = 180 , end angle = 360, radius = 1cm];

  \node[above right] at (45:2cm) {$\Phi$};
\node[above left]  at (135:2cm) {$\Phi$};
\node[above] at (90:1cm) {$S$};
\node[below]  at (270:1cm) {$\chi$};
\node[above] at (180:1.5cm) {$\Psi_a$};
\node[above] at (0:1.5cm) {$\Psi_b$};
\node[left] at (190:2cm) {$u_k$};
\node[below] at (270:2cm) {$u_k$};
\node[right] at (350:2cm) {$u_k$};

  \end{feynman}
\end{tikzpicture}
    \caption{A Feynman diagram contributing to $\mathcal{O}_{W}$.}
    \label{fig:EDM}
\end{figure}
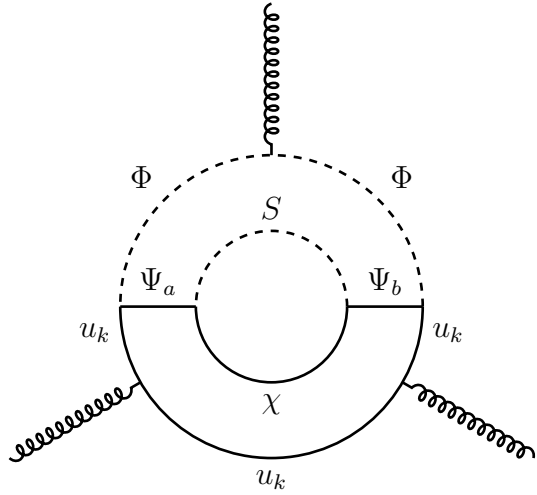

Since we are considering coupling a maximally CPV HS to the SM, we need to consider potentially observable signatures of CPV at low energies. For instance, the SM prediction for the neutron EDM (nEDM) arising from the CKM phase is $d_n^{\rm SM} \simeq 10^{-32} e\, \text{cm}$, which is six orders of magnitude below the current upper bound $d_n < (0.0\pm 1.1) \times 10^{-26} e\, \text{cm}$ \cite{Abel:2020pzs}.
In our model, there are 3-loop contributions to hadron EDM's (see Fig.~\ref{fig:EDM}) 
generating the CPV Weinberg effective operator 
\begin{equation}
    \mathcal{O}_{W} = \frac{1}{3} f^{abc} \epsilon^{\mu \nu \rho \sigma} G_{\mu \lambda}^a G_\nu^{b \lambda} G_{\rho \sigma}^c.
\end{equation}
Parametrically estimating the size of this diagram, we get for the coefficient of $\mathcal{O}_{W}$
\bea
C_G \sim \frac{1}{M_{\Phi}^4} \frac{\Lambda_{\text{QCD}}^2}{(16 \pi^2)^3} g_s^3 \sum_k \Delta_k,
\label{eq:3G}
\eea
where $g_s$ is the QCD coupling evaluated at the scale of the UV physics. $\mathcal{O}_{W}$ contributes to the neutron EDM as~\cite{Weinberg:1989dx}: 
\bea
\Delta d_n = \pm e \Lambda_{\text{nEDM}} C_G,
\eea
where $\Lambda_{\text{nEDM}}  = 10-30$ MeV. Note that for both the top and bottom quark case, the contribution to the neutron EDM is suppressed by $M_\Phi^4$ which is bounded below at $1.2\  \rm TeV$ from collider constraints. As a result, the induced EDMs are significantly below the SM predictions for the entire parameter range of interest, and could only be observed as a small perturbation of the SM value, which appears unlikely.

\section{Conclusions and Outlook}
\label{sec:conclusions}

In this paper we have considered a scenario where the baryon asymmetry in the SM is generated via a portal coupling to a HS, and where there is no explicit breaking of baryon or lepton number. Therefore, no net baryon number is generated, instead out-of-equilibrium decays lead to a sequestering of equal and opposite baryon numbers in the visible and the hidden sectors.

For this mechanism to work, the sequestering of baryon number needs to happen below the electroweak scale such that the asymmetry is not washed out by sphalerons, but still at scales high enough such that predictions of BBN remain unchanged. This leaves only a few orders of magnitude of energy to generate the OBA. Given that the CPV that allows for asymmetry generation arises from the interference between tree and loop level diagrams, this effect is generically suppressed by a loop factor, and the combination of a low efficiency combined with a limited time window in which the asymmetry can be generated presents a potential challenge.

In order to make quantitative statements, we considered the two sectors to be coupled via the baryon portal and two near-degenerate SM singlet fermionic mediators. In a minimal benchmark model, this portal arises from integrating out a color triplet scalar with couplings to right handed up as well as down-type quarks. We also included stable states in the HS to store a baryon number equal and opposite to that of the SM. This minimal scenario nevertheless has room for physical CPV phases, satisfying the second Sakharov condition.

By studying first a case where the asymmetry is generated via the decay of top quarks, we demonstrated that the generic size of CPV is nevertheless sufficient to reach the OBA. This case has the downside that the involved couplings are quite small so this scenario is difficult to test experimentally, unless the new physics that underlies the portal interaction (in our benchmark model, a new colored particle) is within the reach of the LHC or a future collider.

In contrast, if the asymmetry is generated via the decay of bottom quarks, there is an additional small number in play, namely BSM decay channels of bottom quarks are already highly constrained. This, combined with the generically small efficiency of CPV no longer leaves room for generating the OBA, thus the efficiency needs to be enhanced. Using the variant of the mesogenesis mechanism presented in ref.~\cite{Elor:2025fcp}, we have calculated the amount of enhancement that can be obtained through resonant effects between the two nearly mass degenerate portal particles by a combination of analytical and numerical Monte Carlo methods. The main result of our study is that the efficiency $A_{CP}$ for sequestering the baryon number asymmetry between the two sectors can indeed be made $\cO(1)$ without requiring a specific flavor structure in the relevant couplings.

For the mesogenesis scenario, this presents an ultimate challenge. Even with $A_{CP}\sim 1$, a bottom quark branching fraction of at least ${\mathcal O}(10^{-8})$ into the HS is needed for generating the OBA. The current limits are ${\mathcal O}(10^{-5})$, and are expected to improve to ${\mathcal O}(10^{-7})$. Thus the majority, and with additional improvements, all of the parameter space of this scenario may be tested, and the mesogenesis framework can be confirmed or ruled out in full.

For the typical values for the couplings of interest, it appears unlikely that the minimal model can be probed via neutron EDM measurements, however direct searches for the color triplet scalar (or, in other realizations of the baryon portal, a different SM-charged heavy particle) may offer a complementary channel of discovery and confirmation of this scenario for the generation of the baryon asymmetry.

\section*{Acknowledgements}
We thank Gilly Elor. The research of CK and SM is supported by the National Science Foundation under Grant Number PHY-2210562.

\appendix

\section{Calculation of the resonant enhancement}
\label{sec:appCP}

Here we expand on the detailed calculation of the CPV parameter $A_{CP}$ for the HS model introduced in the main text. Recall that CPV arises from the interference between the tree level and 1-loop decays involving dark CPV phases, shown in figures~\ref{fig:psi-production-u} and \ref{fig:psi-production}.
 
\begin{figure}[!h]
    \centering
    
    \begin{tikzpicture}[line width=1pt]
    \begin{feynman}
  \draw[->] (1.1cm, 0.3cm) -- (1.9cm, 0.3cm) node[midway, above=2pt] {$p$};
  \draw (1cm,0) -- node[below] {$\overline{\Psi}_b$} (2cm,0);
  \draw (0,1cm) node[left] {$\Phi$}  -- (1cm,0) ;
  \draw (0,-1cm) node[left] {$u_k$}  -- (1cm,0) ;

  \filldraw[pattern=north east lines, draw=black] (2.5cm,0) circle (0.5cm);
  \draw[-] (3cm,0) -- (4cm,0) node[below] {$\overline{\Psi}_a$};
  \end{feynman}
\end{tikzpicture}
    \caption{The part of the Feynman diagrams which resonantly enhances the asymmetry generation.}
    \label{fig:amp-app}
\end{figure}
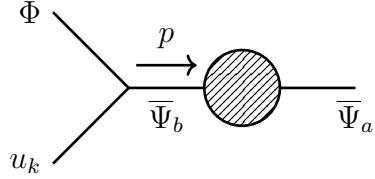

We will be working in the $\Psi$ mass basis with the Dirac fermion decomposed as $\Psi_a = (\psi_a,\overline{\psi_a^c})$ in terms of two left-handed Weyl spinors $\psi_a$ and $\psi_a^c$. Note that everything is common to the tree and loop diagrams except for the leg carrying the momentum $p$. Let us focus our attention on that part of the diagrams, casting it into the form in Fig.\ref{fig:amp-app}. The tree level contribution associated to this subamplitude is
\begin{equation}
     \cM_a^{(\text{tree})}=-i\kappa_{ka}x_{k}\cdot y_{a} ,
\end{equation}
where $x_{k},y_{a}$ are the spinor wavefunctions associated to the external $u$ and $\Psi$ lines, respectively. $x_k$ is to be replaced with $y_k$ when we consider the process where up-type quark is in the final state. For the tree-level diagram, the blob in Fig.~\ref{fig:amp-app} is trivial; for the loop contribution one has $a\neq b$ at the left vertex with a propagator-mixing insertion connecting flavors $b$ and $a$.

 We define the 1-loop contribution to the $\Psi$ propagator as $\Sigma_{ba}$.  Then, in the on-shell renormalization scheme as developed in \cite{Pilaftsis:2003gt} when applied to the diagram in Fig.\ref{fig:amp-app}, we get 
\begin{align}
    \cM_a &=-i \lim_{p^2 \rightarrow m_{\Psi_a}^2 }  x_k\left[  \kappa_{ka} -  \kappa_{kb} \frac{i\slashed{p}}{ p^2 - m_{\Psi_b}^2+im_{\Psi_b}\Gamma_b}\Sigma_{ba}^\text{abs}\right]y_a.
\end{align}
Note that only the absorptive part of $\Sigma$ contributes. Turning our attention to $\Sigma_{ba}^\text{abs}$, we can express
\begin{equation}
    \Sigma_{ba}^\text{abs}=\slashed{p} \sigma_{ba},\qquad (a \neq b)
\end{equation}
where $p$ is the momentum of the external $\Psi$ leg, $ K(x,y,z) = x^2 + y^2 + z^2 -2xy - 2yz - 2zx$ 
is the Källen function, and 
\bea
\sigma_{ba}\equiv \left(-y_b(y_a)^*\frac{1}{8\pi}\frac{1}{p^2}K(p^2,m_\chi^2,m_S^2)^\frac{1}{2}\right).
\eea
When $a=b$, the previous equation can be written as $\sigma_{aa}=\Gamma_a / m_a$, where $\Gamma_a$ is the width of $\Psi_a$. We choose the masses such that the decay $\Psi \to S\,\chi$ kinematically allowed and therefore the loop has an imaginary part.

Combining everything,
\begin{align}
    \cM_a=-ix_k\left[  \kappa_{ka} - i  \kappa_{kb} \frac{\sigma_{ba}\left(1+\frac{m_{\Psi_a}}{m_{\Psi_b}}(1+i\frac{\Gamma_b}{m_{\Psi_b}})\right)}{R-\frac{m_{\Psi_a}}{m_{\Psi_b}}\left(\frac{\Gamma_b}{m_{\Psi_b}}\right)^2+2i\frac{\Gamma_b}{m_{\Psi_b}}\frac{m_{\Psi_a}}{m_{\Psi_b}}}\right] y_a,
\end{align}
where we have used the on-shell condition for the external spinor and we have defined
\begin{equation}
    R\equiv \frac{m_{\Psi_b}^2-m_{\Psi_a}^2}{m_{\Psi_b}m_{\Psi_a}}.
\end{equation}
Simplifying this, we can cast the amplitude into the form
\begin{align}
    \cM_a &\approx-ix_k\cdot y_a\left[ \kappa_{ka} - i  \kappa_{kb} \frac{\sigma_{ba}\left(2+i\frac{\Gamma_b}{m_{\Psi_b}})\right)}{R+2i\frac{\Gamma_b}{m_{\Psi_b}}}\right]\\
    &= -ix_k \cdot y_a\frac{1}{D_b}\left[  \kappa_{ka}D_b - i  \kappa_{kb}{\sigma_{kb}\left(2+i\frac{\Gamma_b}{m_{\Psi_b}}\right)\left(R-2i\frac{\Gamma_b}{m_{\Psi_b}}\right)}\right],
\end{align}
where we defined $D_b\equiv R^2+4 \left(\frac{\Gamma_b}{m_{\Psi_b}}\right)^2$.
The quantity of interest is 
\bea
A_{CP}^{a} \equiv
 \frac{\Sigma_{\text{spins}}\int d\text{LIPS}_3(|\cM|^2-|\overline{\cM}|^2)}{\Sigma_{\text{spins}}\int d\text{LIPS}_3(|\cM|^2+|{\overline{\cM}}|^2)},
\eea
  where $\int d\text{LIPS}_3$ is the Lorentz invariant phase space integral for three final state particles, and $\cM$ is the full amplitude shown in Figs.\ref{fig:psi-production-u} and \ref{fig:psi-production} with $\Psi^a$ in the final state. $\overline{\cM}$ is the amplitude for the CP conjugate process, at the same point in phase space. For our model, (and more generally, with the caveat of considering vertex corrections as subdominant~\cite{Pilaftsis:2003gt}), the phase space dependence of the numerator and denominator cancel out, leaving just a ratio involving combinations of couplings. 

We are now in a position to calculate the amount of CPV. We arrive at the compact formula, 
\begin{align}
    A_{CP}^{a} &\approx  \frac{4\text{Im} (-\kappa_{ka}\kappa_{kb}^*\sigma_{ba}^*)\times D_b\times 2R}{|\kappa_{ka}|^2|D_b^2}\\
    &= \frac{\text{Im} (\kappa_{ka}\kappa_{kb}^*(y_b)^*y_a)}{2\pi m_{\Psi_a}^2|\kappa_{ka}|^2}\frac{R}{R^2+4\left(\frac{\Gamma_b}{m_{\Psi_b}}\right)^2}K (m_{\Psi_a}^2,m_\chi^2,m_S^2)^\frac{1}{2}.
    \label{eq:appfinalresult}
\end{align}
Note that if we had set $\Gamma=0$ from the start, the RHS would diverge in the degenerate limit as $\frac{1}{\Delta m_{\Psi}}$. Finite width effects smooth out this divergence. For a rough estimate, we can set $K (m_{\Psi_a}^2,m_\chi^2,m_S^2)^\frac{1}{2}\approx m_{\Psi_a}^2$. Furthermore, for simplicity, one can approximate $\Gamma_a \approx \Gamma_b \equiv \Gamma$. With these approximations, our formula takes on the follwowing simplified form
\begin{align}
    A_{CP}^{1}&\approx\frac{\Delta_k}{4\pi |\kappa_{k1}|^2m_{\Psi}}\frac{\Delta m_{\Psi}K(m_{\Psi}^2,m_\chi^2,m_S^2)^\frac{1}{2}}{(\Delta m_{\Psi})^2+\Gamma^2}\\
    &\approx \frac{\Delta_k}{4\pi |\kappa_{k1}|^2}\frac{m_{\Psi}\Delta m_{\Psi}}{(\Delta m_{\Psi})^2+\Gamma^2}.
    \label{res}
\end{align}
where $\Delta m_{\Psi}=m_{\Psi_1}-m_{\Psi_2}$, $\Delta_k = \text{Im} (\kappa_{k1}\kappa_{k2}^*(y_2)^*y_1)$ and $m_{\Psi}\approx m_{\Psi_1}\approx m_{\Psi_{2}}$. Note that this expression agrees with the result of ref.~\cite{Garny:2011hg}. It is straightforward to observe that the CPV is most efficient when $\Gamma\sim m_{\Psi}$. 

\makeatletter
\renewcommand{\@seccntformat}[1]{Appendix \csname the#1\endcsname\quad}
\makeatother
\renewcommand{\thesection}{\Alph{section}}
\numberwithin{equation}{section}

\clearpage
\bibliographystyle{JHEP}
\bibliography{ref}{}

\end{document}